\documentclass[nomathfonts]{aipproc} 
\layoutstyle{6s}

\title{On the estimation of a parameter with incomplete knowledge on a nuisance parameter
}
\author{Ali Mohammad-Djafari}
 {
  address = {Laboratoire des Signaux et Syst\`emes,\linebreak 
  Unit\'e mixte de recherche 8506 (CNRS-Sup\'elec-UPS) \linebreak  
  Sup\'elec, Plateau de Moulon, 91192 Gif-sur-Yvette, France},
  email = {djafari@lss.supelec.fr}
 }
\author{Adel Mohammadpour}
{
 address={School of intelligent Systems, IPM, Tehran, Iran,\linebreak 
 and \linebreak 
 {Department of Statistics, Faculty of Mathematics \& Computer  Science, Amirkabir University of Technology, Tehran, Iran}\linebreak },
 email={\tt adel@aut.ac.ir}
 }
\date{}

\def\beq{\begin{equation}}
\def\eeq{\end{equation}}
\def\bit{\begin{itemize}}
\def\eit{\end{itemize}}
\def\beqn{\begin{eqnarray}}
\def\eeqn{\end{eqnarray}}
\def\beqnx{\begin{eqnarray*}}
\def\eeqnx{\end{eqnarray*}}

\def\d#1{\;\mbox{d}#1}
\def\esp#1{\mbox{E}\left\{#1\right\}}
\def\espx#1#2{\mbox{E}_{#1}\left\{#2\right\}}
\def\med#1{\mbox{Median}\left\{#1\right\}}
\def\expf#1{\mbox{exp}\left\{#1\right\}}
\def\argmax#1#2{\displaystyle{\mbox{arg}\max_{#1}\left\{#2\right\}}}
\def\argmin#1#2{\displaystyle{\mbox{arg}\min_{#1}\left\{#2\right\}}}

\def\Nc{{\cal N}}

\begin{document}

\def\thetah{\widehat{\theta}}
\def\sigmah{\widehat{\sigma}}

\def\Ftxm{\widetilde{F}_{X}(x_-)}
\def\Ftxp{\widetilde{F}_{X}(x_+)}

\def\NU{{\cal V}}

\def\tFx{\widetilde{F}_{X}}

\def\Fx{F_{X}(x)}
\def\fx{f_{X}(x)}

\def\Fxv{F_{X}(x)}
\def\fxv{f_{X}(x)}

\def\Fxi{{F}_{X_i}(x_i)}
\def\fxi{{f}_{X_i}(x_i)}

\def\Gnu{G_{\NU}(\nu)}
\def\gnu{g_{\NU}(\nu)}

\def\Fnu{F_{\NU}(\nu)}
\def\fnu{f_{\NU}(\nu)}

\def\Finnu#1{F^{-1}_{\NU}(#1)}

\def\Gnua#1{G_{\NU}(#1)}
\def\Gnub#1{G_{\NU}^{-1}(#1)}
\def\Gnuh{G_{\NU}^{-1}(1/2)}

\def\Ftx{\widetilde{F}_{X}(x)}
\def\ftx{\widetilde{f}_{X}(x)}

\def\Ftxi{\widetilde{F}_{X_i}(x_i)}
\def\ftxi{\widetilde{f}_{X_i}(x_i)}

\def\Ftxv{\widetilde{F}_{X}(x)}
\def\FtX#1{\widetilde{F}_{X}(#1)}
\def\ftxv{\widetilde{f}_{X}(x)}

\def\fxnu{f_{X|\NU}(x|\nu)}
\def\fxNU{f_{X|\NU}(x|\NU)}
\def\FxNU{F_{X|\NU}(x|\NU)}

\def\fxNUtheta{f_{X|\NU,\theta}(x|\NU,\theta)}
\def\FxNUtheta{F_{X|\NU,\theta}(x|\NU,\theta)}

\def\Fxnu{F_{X|\NU}(x|\nu)}
\def\Fxnui#1{F_{X|\NU}(x|\nu_#1)}

\def\FxN#1{F_{X|\NU}(#1)}
\def\FxNU{F_{X|\NU}(x|\NU)}
\def\Fxnun#1{F_{X|\NU}(x|#1)}
\def\Fxinun#1{F_{X_i|\NU}(x_i|#1)}
\def\fxnu{f_{X|\NU}(x|\nu)}
\def\fxNU{f_{X|\NU}(x|\NU)}

\def\Fxvnu{F_{X|\NU}(x|\nu)}
\def\Fxvnun#1{F_{X|\NU}(x|#1)}
\def\FxvNU{F_{X|\NU}(x|\NU)}

\def\FXvnu{F_{X|\NU}}

\def\fxvnu{f_{X|\NU}(x|\nu)}
\def\fxvNU{f_{X|\NU}(x|\NU)}

\def\FXvNU#1{{F}_{X|\NU}(#1|\NU)}
\def\FXvn#1{{F}_{X|\NU}(x|#1)}

\def\Fxtheta{F_{X|\theta}(x|\theta)}
\def\fxtheta{f_{X|\theta}(x|\theta)}

\def\Fxvtheta{F_{X|\theta}(x|\theta)}
\def\fxvtheta{f_{X|\theta}(x|\theta)}

\def\Ftxtheta{\widetilde{F}_{X|\theta}(x|\theta)}
\def\ftxtheta{\widetilde{f}_{X|\theta}(x|\theta)}

\def\Ftxvtheta{\widetilde{F}_{X|\theta}(x|\theta)}
\def\ftxvtheta{\widetilde{f}_{X|\theta}(x|\theta)}

\def\Fxnutheta{F_{X|\NU,\theta}(x|\nu,\theta)}
\def\fxnutheta{f_{X|\NU,\theta}(x|\nu,\theta)}

\def\ftheta{f_{\Theta}(\theta)}
\def\fthetaxnuz{f_{\Theta|X,\nu_0}(\theta|x,\nu_0)}
\def\fthetax{f_{\Theta|X}(\theta|x)}

\def\Fxvnutheta{F_{X|\NU,\theta}(x|\nu,\theta)}
\def\FxvNUtheta{F_{X|\NU,\theta}(x|\NU,\theta)}
\def\fxvnutheta{f_{X|\NU,\theta}(x|\nu,\theta)}
\def\Ftxvnutheta{\widetilde{F}_{X|\theta}(x|\theta)}
\def\ftxvnutheta{\widetilde{f}_{X|\theta}(x|\theta)}

\def\FxvNUm{F_{X|\NU}(x_-|\NU)}
\def\FxvNUp{F_{X|\NU}(x_+|\NU)}

\def\Fxinu{F_{X_i|\NU}(x_i|\nu)}
\def\fxinu{f_{X_i|\NU}(x_i|\nu)}

\def\Ftxi{\widetilde{F}_{X_i}(x_i)}
\def\ftxi{\widetilde{f}_{X_i}(x_i)}

\def\fxitheta{f_{X_i|\theta}(x_i|\theta)}
\def\ftxitheta{\widetilde{f}_{X_i|\theta}(x_i|\theta)}

\def\thetah{\widehat{\theta}}
\def\thetat{\widetilde{\theta}}

\def\lnuxv{l_{\nu|X}(\nu|x)}

\def\Lnuxva#1{L_{\nu|X}(#1 | x)}
\def\lnuxva#1{l_{\nu|X}(#1 | x)}

\def\Prob#1{\mbox{P}\left(#1\right)}

\def\lthetaxvnuz{l(\theta)}

\def\lthetaxvnuz{l_{\theta|x,\nu_0}(\theta | x,\nu_0)}
\def\fxnuztheta{f_{X|\nu_0,\theta}(x | \nu_0,\theta)}
\def\fxinuztheta{f_{X|\nu_0,\theta}(x_i | \nu_0,\theta)}
\def\fxvtheta{f_{X|\theta}(x | \theta)}
\def\lthetaxv{l_{\theta|X}(\theta | x)}
\def\ltthetaxv{\widetilde{l}_{\theta|X}(\theta | x)}

\def\ltheta{l(\theta )}
\def\lttheta{\widetilde{l}(\theta)}
\def\Lnu{L(\nu )}
\def\Lnui#1{L(\nu_#1)}
\def\Linu{L_i(\nu )}
\def\Linu{L_i(\nu )}
\def\LNU{L(\NU )}
\def\Lin#1{L^{-1}(#1)}
\def\L#1{L(#1)}
\def\Li#1{L_i(#1)}
\def\FNU#1{F_\NU(#1)}
\def\FinNU#1{F^{-1}_\NU(#1)}

\def\ftthetax{\tilde{f}_{\Theta|X}(\theta|x)}
\def\ER{R}

\pagenumbering{arabic} \setcounter{page}{1}

\begin{abstract}
In this paper we consider the problem of estimating a parameter of a probability distribution when we have some prior information on a nuisance parameter. 
We start by the very simple case where we know perfectly the value of the nuisance parameter. The complete likelihood is the classical tool in this case.  Then, progressively, we consider the case where we are given a prior probability distribution on this nuisance parameter. The marginal likelihood is then the  classical tool in this case.  Then, we consider the case where we only have a fixed number of its moments. Here, we may use the maximum entropy (ME) principle to assign a prior law and thus go back to the previous case. Finally, we consider the case where we know only its median. In our knowledge, there is not any classical tool for this case. We propose then a new tool for this case based on a recently proposed alternative distribution to the marginal probability distribution. 
This new criterion is obtained by first remarking that the marginal distribution can be considered as the mean value of the original distribution over the prior probability law of the nuisance parameter, and then, by using the median in place of the mean. 
In this paper, we first summarize the classical tools used for the three first cases, then we give the precise definition of this new criterion and its properties and, finally, present a few examples to show the differences of these cases. 
\\ ~\\ 
{\bf Key Words: } 
Nuisance parameter, Bayesian inference, Maximum Entropy, Marginalization,  Incomplete knowledge, Mean and Median of the Likelihood over the prior distribution
\end{abstract}

\maketitle

\section{Introduction}
We consider the problem of estimating a parameter of interest $\theta$ of a probability distribution when we have some prior information on a nuisance parameter $\nu$ from only one or a finite number of samples from this probability distribution. Assume that we know the expression of either 
the cumulative distribution function (cdf) 
$\Fxnutheta$ or equivalently the expression of its probability density function (pdf) $\fxnutheta$. 
We assume that $\nu$ is a nuisance parameter on which we have an 
\emph{a priori} information. This prior information can either be complete knowledge of its value $\nu_0$ or more and more incomplete such as a prior distribution $\Fnu$ (or a pdf $\fnu$) or only the knowledge of a finite number of its moments or still just the knowledge of its median. For the three first  cases there are classical solutions, but in our knowledge, there is not yet any solution for this last case. The main object of this paper is to propose a solution for it. This solution is based on a recently proposed inference tool which is obtained using the median in place of 
the mean when using a prior distribution on the nuisance parameter 
\cite{Mohammadpour03a,Mohammadpour03b,Rohatgi76,Mohammadpour04a}.

This paper is then organized as follows. First, we give a brief presentation of the three well known approaches. Then, we summarize the recently proposed inference tool and we will see how we can use it for the last problem. 
 
\section{Classical approaches of parameter estimation}
Assume that we are given an observation $x$ and assume that its cumulative distribution function (cdf)  
$\Fxnutheta$ (or equivalently its probability density function (pdf)  $\fxnutheta$) depends on two parameters $\nu$ and $\theta$. We assume that $\theta$ is the parameter of interest  and 
$\nu$ is a nuisance parameter. 
We are looking for tools to infer $\theta$ from one observation $x$ and some prior knowledge on $\nu$. We are then going to consider the following cases: 

\bigskip\noindent{\bf 
Perfect knowledge of $\nu$, i.e., $\nu=\nu_0$:}\\ ~\\ 
Then, the classical approach is the Maximum Likelihood (ML) estimate 
\beq
\thetah_{ML}=\argmax{\theta}{l_0(\theta)=\fxnuztheta}.
\label{ML}
\eeq
If we also have a prior $\ftheta$ on the parameter of interest $\theta$, then we can use the Bayesian approach by computing the a posteriori distribution $\fthetaxnuz$ and then use any estimator such as the Maximum {\it a posteriori} (MAP) estimate 
\beq
\thetah_{MAP}=\argmax{\theta}{\fthetaxnuz}
=\argmax{\theta}{l_0(\theta) \, \ftheta} 
\label{MAP}
\eeq
or the Bayesian Mean Square Estimate (MSE) 
\beq
\thetah_{MSE}=\esp{\Theta}
={\int \theta \, \fthetaxnuz \d{\theta}}
=\frac{\int \theta \, l_0(\theta) \, \ftheta \d{\theta}}
{\int l_0(\theta) \, \ftheta \d{\theta}}.
\label{MSE}
\eeq

\bigskip\noindent{\bf 
Incomplete knowledge of $\nu$ through an apriori cdf $\Fnu$ or pdf $\fnu$:}\\ ~\\ 
The classical approach here is the Marginal Maximum Likelihood (MML) estimate 
\beq
\thetah_{MML}=\argmax{\theta}{l_1(\theta)=\fxtheta} 
\label{MML}
\eeq
where 
\beq
\fxtheta=\int \fxnutheta \fnu \d{\nu} .
\eeq
Again here, if we also have a prior $\ftheta$ we can define the \emph{a posteriori} distribution $\fthetax$ and 
\beq
\thetah_{MMAP}=\argmax{\theta}{\fthetax}
=\argmax{\theta}{l_1(\theta) \, \ftheta}
\label{MMAP}
\eeq
or
\beq
\thetah_{MMSE}=\esp{\Theta}={\int \theta \, \fthetax \d{\theta}}
=\frac{\int \theta \, l_1(\theta) \, \ftheta \d{\theta}}
{\int l_1(\theta) \, \ftheta \d{\theta}} .
\label{MMSE}
\eeq

\bigskip\noindent{\bf 
Incomplete knowledge of $\nu$ through the knowledge of a finite number of its moments:} \\ ~\\ 
Assume now that our prior knowledge on the parameter $\nu$ is expressed through the knowledge of a finite number of the moments:
\beq
\esp{\phi_k(\NU)}=d_k, \quad k=1,\cdots,K
\eeq
where $\phi_k$ are known functions. Particular cases are $\phi_k(\nu)=\nu^k$ where $\{d_k, \; k=1,\cdots,K\}$ are then the moments up to order $K$ of $\NU$. 

Here, we can use the principle of Maximum Entropy (ME) to assign a prior probability law $\ftheta$ which is the classical tool for assigning a probability law to a quantity when we know only a finite number of its moments. The solution is well known and is given by 
\beq
\fnu=\expf{-\lambda_0 -\sum_{k=1}^K \lambda_k \phi_k(\nu)}
\eeq
where the Lagrange parameters $\{\lambda_k, \; k=0,\cdots,K\}$ are the solution of the following system of equations:
\beq
\int \phi_k(\nu) \; \expf{-\lambda_0 -\sum_{k=1}^K \lambda_k \phi_k(\nu)} \d{\nu} = d_k, \quad k=1,\cdots,K, 
\eeq
where we used $\phi_0(\nu)=1$ and $d_0=1$ to include the normalization factor $\lambda_0$. 
For more details on ME and also on the computational aspects of the Lagrange parameters refer to \cite{Shannon48,Jaynes68,Verdugo78,Agmon79,Shore80,Jaynes82,Shore83,Mukherjee84,Titterington84a,Djafari90,Djafari91,Djafari91a,Borwein91a,Borwein91b}. 

From this point, i.e., when we obtain an expression for $\fnu$ which translates our prior knowledge of the moments on the nuisance parameter, 
the problem becomes equivalent to the previous case. 

\bigskip\noindent{\bf 
Incomplete knowledge of $\nu$ through the only knowledge of its median  
\emph{(New alternative criterion)}:} \\ ~\\ 
Assume now that our prior knowledge on the parameter $\nu$ is expressed through the knowledge of its median value. Up to our knowledge, we do not have any classical tool such as ME of the previous case to translate this knowledge into a probability law $\ftheta$. The main contribution of this paper is exactly to provide a coherent solution to this case which is detailed in the next section. 

\section{New inference tool}

Recently, we proposed a new alternative to the classical approach for this case which consists in proposing an alternative criterion $\ftxtheta$ (or equivalently 
$\Ftxtheta$) to the \emph{likelihood function} $\fxtheta$ (or equivalently 
$\Fxtheta$) which we called \emph{likelihood based on the median} which can be used in place of $\fxtheta$ in the previous case. 

The name \emph{likelihood based on the median} for $\ftxtheta$ is motivated by the fact that $\fxtheta$ in 
\beq
\fxtheta=\int \fxnutheta \fnu \d{\nu}=\espx{\NU}{\fxNUtheta}, 
\eeq
or equivalently $\Fxtheta$ in 
\beq
\Fxtheta=\int \Fxnutheta \fnu \d{\nu}=\espx{\NU}{\FxNUtheta},
\eeq
can be recognized as the mean value of $\fxNUtheta$ (or $\FxNUtheta$) over the probability law $\fnu$. 

The proposed new criterion is then defined as the median value of $\fxNUtheta$ (or $\FxNUtheta$) over the probability law $\fnu$: 
\beqnx
\Ftxvtheta &:& \Prob{\FxvNUtheta \le \Ftxvtheta}=1/2
\eeqnx

In previous works, we showed that, under some mild conditions on $\Fxnutheta$ 
the function $\Ftxtheta$ (strictly increasing) has all the properties of a cdf 
and thus the function $\tilde{l}_1(\theta)=\ftxtheta$ 
has all the properties of a likelihood function. 
Thus, we can use it in place of $l_1(\theta)$, i.e.:

\beq
\thetah_{MLM}=\argmax{\theta}{\tilde{l}_1(\theta)=\ftxtheta}
\label{MLM}
\eeq
or if we also have a prior $\ftheta$ 
\beq
\thetah_{MAPM}=\argmax{\theta}{\ftthetax}
=\argmax{\theta}{\tilde{l}_1(\theta) \, \ftheta}
\label{MAPM}
\eeq
or
\beq
\thetah_{MSEM}=\esp{\Theta}={\int \theta \, \ftthetax \d{\theta}}
=\frac{\int \theta \, \tilde{l}_1(\theta) \, \ftheta \d{\theta}}{\int \tilde{l}_1(\theta) \, \ftheta \d{\theta}} . 
\label{MSEM}
\eeq

Indeed, we showed that the expression of $\Ftxtheta$ is given by
\beq
\Ftxtheta =L\left( \Finnu{\frac{1}{2}} \right)
\eeq
where $\Lnu =\Fxnutheta$. 
Thus to obtain the expression of $\Ftxtheta$ we only need to know the median 
value $\Finnu{\frac{1}{2}}$ of the distribution $\fnu$. 

In what follows, we are considering the four aforementioned cases, i.e. 
i) the perfect knowledge $\nu=\nu_0$, 
ii) the knowledge of $\fnu$, 
iii) the knowledge of the mean value $\bar{\nu}$ of $\NU$ and 
iv) the knowledge of the median $\tilde{\nu}$ of $\NU$, 
and examine them through a simple but difficult case where we have only one observation 
$x$ of $X$ with the pdf $\fxnutheta$ and where we want to estimate $\theta$ with the aforementioned knowledge on the nuisance parameter $\nu$. 

\section{Examples}
In what follows, we use the following notations and expressions: 

\def\Ndd#1#2#3{{\cal N}\left( #1 ;~ #2, #3 \right)}
\def\Nddx#1#2#3{(2\pi{#3})^{-\frac{1}{2}} \expf{-\frac{1}{2{#3}}(#1-#2)^2}}
\def\NddL#1#2#3{\frac{1}{2}\ln #3 + \frac{1}{2 #3}(#1-#2)^2}

\def\Edd#1#2{{\cal E}\left( #1 ;~ #2 \right)}
\def\Eddx#1#2{#2 \expf{-#1 / #2}}

\def\DEdd#1#2{{\cal DE}\left( #1 ;~ #2 \right)}
\def\DEddx#1#2{\frac{#2}{2} \expf{- |#1| / #2}}

\def\Gdd#1#2#3{{\cal G}\left( #1 ;~ #2, #3 \right)}
\def\Gddx#1#2#3{\frac{#3^{#2}}{\Gamma(#2)} x^{#2-1}\expf{-#3 #1}}

\def\IGdd#1#2#3{{\cal IG}\left( #1 ;~ #2, #3 \right)}
\def\IGddx#1#2#3{\frac{#3^{#2}}{\Gamma(#2)} #1^{#2+1}\expf{-#3/#1}}

\def\Cdd#1#2#3{{\cal C}\left( #1 ;~ #2, #3 \right)}
\def\Cddx#1#2#3{4\pi\left(1+\frac{#1-#2}{#3}\right)^{\frac{-1}{2}}}

\def\Sdd#1#2#3#4{{\cal S}\left( #1 ;~ #2, #3, #4 \right)}
\def\Sddx#1#2#3#4{\frac{\Gamma((#4+1)/2)}{\Gamma((#4)/2)(\theta\pi)^{1/2}} \left(1+\frac{1}{#4}\frac{#1-#2}{\theta}\right)^{\frac{-(#4+1)}{2}}}

\[
\begin{array}{llcl}
\mbox{Gaussian:} &
\Ndd{x}{\mu}{\sigma^2}
&=& \Nddx{x}{\mu}{\sigma^2}
\\ 
\mbox{Exponential:} &
\Edd{x}{\lambda}
&=&  \Eddx{x}{\lambda}
\\ 
\mbox{Double Exponential:} &
\DEdd{x}{\lambda}
&=& \DEddx{x}{\lambda}
\\
\mbox{Gamma:} &
\Gdd{x}{\alpha}{\beta}
&=&\Gddx{x}{\alpha}{\beta}
\\
\mbox{Inverse Gamma:} &
\IGdd{x}{\alpha}{\beta}
&=& \IGddx{x}{\alpha}{\beta}
\\
\mbox{Student:} &
\Sdd{x}{\mu}{\theta}{\alpha}
&=& \Sddx{x}{\mu}{\theta}{\alpha}
\\
\mbox{Cauchy:} &
\Cdd{x}{\mu}{\theta}
&=&4\pi\left(1+\frac{x-\mu}{\theta}\right)^{\frac{-1}{2}}  
\end{array}
\]

\subsection{Example 1}
The first example we consider is 
\[
\fxnutheta=\Ndd{x}{\nu}{\theta}=\Nddx{x}{\nu}{\theta}
\] 
where we assume that the mean value $\nu$ is the nuisance parameter. Then: 
\bit
\item  Complete knowledge case $\nu=\nu_0$: \\ 
Then we have
\[
\fxnuztheta=\Ndd{x}{\nu_0}{\theta}=\Nddx{x}{\nu_0}{\theta}
\]
and the ML estimate of $\theta$ is obtained by 
\[
\thetah=\argmax{\theta}{\fxnuztheta}=\argmin{\theta}{L(\theta)=\NddL{x}{\nu_0}{\theta}}
\] 
which gives $\thetah=(x-\nu_0)^2$. 

\item Prior pdf case $\fnu=\Ndd{\nu}{\nu_0}{\theta_0}$: \\ 
Then we have
\beqnx
\fxtheta
&=&\int \fxnutheta \; \fnu \d{\nu}\\ 
&=&\int \Ndd{x}{\nu}{\theta} \, \Ndd{\nu}{\nu_0}{\theta_0} \d{\nu}\\ 
&=&\int \Nddx{x}{\nu}{\theta} \, \Nddx{\nu}{\nu_0}{\theta_0} \d{\nu}
\eeqnx
and it is not difficult to show that 
$\fxtheta=\Ndd{x}{\nu_0}{\theta+\theta_0}$ and the MML estimate 
of $\theta$ is obtained by 
\[
\thetah=\argmax{\theta}{\fxtheta}
\]
which gives $\thetah=\max((x-\nu_0)^2-\theta_0,0)$. 

\item Moments knowledge case $\esp{\NU}=\nu_0$:\\  
Then, we need also to know the support ${\cal S}$ 
of $\nu$ to be able to use ME and assign $\fnu$. 
If ${\cal S}=\ER$, the ME pdf does not exist, but if ${\cal S}=\ER^+$, 
the ME pdf $\fnu$ is an exponential $\Edd{\nu}{\nu_0}$. 
In this case, we cannot obtain an analytical expression for $\fxtheta$
\beqnx
\fxtheta
&=&\int \Ndd{x}{\nu}{\theta} \, \Edd{\nu}{\nu_0} \d{\nu}\\
&=&\int_0^{\infty} \Nddx{x}{\nu}{\theta} \, \Eddx{\nu}{\nu_0} \d{\nu}
\eeqnx
However, the MML estimate can be computed numerically. 

We may also note that, if we are given $\esp{|\NU|}=\nu_0$, then even for the case ${\cal S}=\ER$ the ME pdf exists and is given by $\DEdd{\nu}{\nu_0}$. 
In this case we have
\beqnx
\fxtheta
&=&\int \Ndd{x}{\nu}{\theta} \, \DEdd{\nu}{\nu_0} \d{\nu}\\
&=&\int \Nddx{x}{\nu}{\theta} \, \DEddx{\nu}{\nu_0} \d{\nu}
\eeqnx
We cannot obtain an analytical expression for $\fxtheta$, 
but again the MML estimate can be computed numerically. 

Finally, if we are given $\esp{\NU}=\nu_0$ and $\esp{(\NU-\nu_0)^2}=\theta_0$, 
then the ME pdf is the Gaussian $\Ndd{\nu}{\nu_0}{\theta_0}$ and 
we can go back to the case of previous item. 

\item Median knowledge case $\med{\NU}=\nu_0$:\\  
Then, as we could see, we have 
$\ftxtheta=\Ndd{x}{\nu_0}{\theta}$ and we can estimate $\theta$ by 
\[
\thetah=\argmax{\theta}{\ftxtheta}
\]
which gives $\thetah=(x-\nu_0)^2$. 
\eit

\subsection{Example 2}
The second example we consider is \[
\fxnutheta=\Ndd{x}{\theta}{\nu}=\Nddx{x}{\theta}{\nu}
\] 
where, this time, we assume that $\nu$ is the variance and the nuisance parameter. Then: 
\bit
\item  Complete knowledge case $\nu=\nu_0$:\\ 
The ML estimate of $\theta$ is obtained by 
$\thetah=\argmax{\theta}{\fxnuztheta}$ which gives $\thetah=x$. 

\item Prior pdf case $\fnu=\IGdd{\nu}{\alpha/2}{\beta/2}$: \\ 
Then, 
\beqnx
\fxnutheta
&=&\int \Ndd{x}{\theta}{\nu} \, \IGdd{\nu}{\alpha/2}{\beta/2} \d{\nu}\\
&=&\int \Nddx{x}{\theta}{\nu} \, \IGddx{\nu}{\alpha/2}{\beta/2} \d{\nu}\\
&=&\Sdd{x}{\theta}{\alpha/\beta}{\alpha}
\eeqnx
and we can estimate $\theta$ by 
$\thetah=\argmax{\theta}{\fxtheta}$ which gives $\thetah=x$. 

\item Moments knowledge case $\esp{\NU}=\nu_0$:\\  
Then, knowing that the variance is a positive quantity $({\cal S}=\ER^+)$, 
the ME pdf $\fnu$ is an exponential $\Edd{\nu}{\nu_0}$. 
In this case we have
\beqnx
\fxtheta
&=&\int \Ndd{x}{\theta}{\nu} \, \Edd{\nu}{\nu_0} \d{\nu}\\ 
&=&\int \Nddx{x}{\theta}{\nu} \, \Eddx{\nu}{\nu_0} \d{\nu}\\ 
&=&\Sdd{x}{\theta}{0}{1}=\Cdd{x}{\theta}{1}
\eeqnx
and $\thetah=x$. 

\item Median knowledge case $\med{\NU}=\nu_0$:\\  
Then, as we could see, we have 
$\ftxtheta=\Ndd{x}{\theta}{\nu_0}$ and we can estimate $\theta$ by 
$\thetah=\argmax{\theta}{\ftxtheta}$ which gives 
$\thetah=x$. 

\medskip 
We may note that, all the estimations of the mean $\theta$ when the nuisance parameter $\nu$ is the variance do not depend on the knowledge of this variance. The reason is that all the likelihood based estimators of a position parameter are scale invariant. 
\eit

\section{Conclusions}
In this paper we considered the problem of estimating one of the two parameters of a probability distribution when the other one is considered as a nuisance parameter on which we may have some prior information. 
We then considered and compared four cases: \\
i) the complete knowledge case where the nuisance parameter is known exactly. This is the simplest case and the classical likelihood based methods apply. \\
ii) the incomplete knowledge case where our prior knowledge is translated through a prior probability distribution. In this case, we can integrate out the nuisance parameter and obtain a marginal likelihood and use it for estimating the parameter of interest. \\
iii) the incomplete knowledge case where our prior knowledge is given to us in the form of a finite number of its moments. In this case, we can use the ME principle to translate our prior knowledge into a prior pdf and find the situation of the previous case. \\
iv) the incomplete knowledge case where our prior knowledge is only the median value of the nuisance parameter. For this case, up to the knowledge of the authors, there is not any classical approach and based on our previous works, we presented a new inference tool which can handle this case. 

Finally, to illustrate these cases, we presented a few examples to show the similarities and differences of these cases.

\def\sca#1{{\sc #1}}
\bibliographystyle{ieeetr}
\bibliography{bibenabr,revuedef,revueabr,baseAJ,baseKZ,gpipubli,biblio}

\end{document}